\documentclass[twocolumn,showpacs,amsmath,amssymb,prd,aps]{revtex4}
\usepackage{graphicx}
\usepackage{amsfonts}
\usepackage{amssymb}
\usepackage{amsmath}
\usepackage{flafter}
\usepackage{epstopdf}
\usepackage{hyperref}
\usepackage{xcolor}
\usepackage{float}
\usepackage[stable]{footmisc}

\begin{document}

\title{A Novel Radiation to Test Foundations of Quantum Mechanics\footnote{This paper was originally posted as SLAC-PUB-7105 in 1996.}}

\date{\today} 
\author{Pisin Chen}
\affiliation{SLAC National Accelerator Laboratory, Menlo Park, CA 94025, USA}
\email{pisinchen@phys.ntu.edu.tw}

\addtocounter{footnote}{-20}

\begin{abstract}
We point out that a new mechanism for radiation should exist if the Bohm theory of quantum mechanics is taken seriously. By traversing a quantum potential, an electron will necessarily be accelerated and radiate. For an illustration, we show that in the double-slit experiment this radiation yields a characteristic spectrum and a distinct pattern on the screen that is complementary to the pattern of the electrons. Experimentally, either the existence or the nonexistence of such a radiation would have important implications for the foundations of quantum mechanics.
\vspace{3mm}


\end{abstract}

\maketitle

Quantum mechanics is hailed as one of the most successful theories in the entire history of physics. It has been applied to a very wide range of phenomena with extremely great accuracy, and there is as yet no experimental indication of any domain where it might break down. Nevertheless, certain notions of quantum mechanics in its foundation continue to be a subject of confusion since the celebrated Bohr-Einstein dialogue.

As is well-known, in the standard Copenhagen interpretation one is not able (or allowed) to describe the motion (i.e., the momentum and position simultaneously) of a particle between measurements. In 1952 David Bohm put forward a causal interpretation \cite{Bohm1952} of quantum mechanics. In the Bohm theory, a particle, say an electron, plays a dual role: on the one hand it has a localized body where its mass and other internal structures, such as charge and spin, reside; on the other hand the particle is involved, together with its environment, in giving rise to a wave function of the whole system, which satisfies Schr\"{o}dingerÕs equation and which is never created nor destroyed. The particle is in turn guided by the so-called quantum potential induced by the wave function. Therefore, in this theory a particle has a well-described trajectory, while the uncertainty principle is obeyed at the epistemological level because of the random initial conditions of the physical process under consideration. It has been argued that the two interpretations yield identical observational results where the variable which describes the motion of the particle is hidden, and thus the difference is only ontological \cite{Bohm1987,BohmHiley}

In an attempt to bring the earlier metaphysical discussions back to a physical plane, in this letter we point out that an electron which is deflected (accelerated) by the quantum potential should necessarily radiate, if the Bohm theory is to be consistent with the Maxwell theory of electrodynamics. As an illustration, we show that in the double-slit experiment this radiation produces a well-described power spectrum and a distinct spatial pattern which compliments the interference pattern of the electrons.

Either the existence or the non-existence of such a radiation should have a profound impact on our understanding of the foundations of quantum mechanics: should this radiation be unambiguously detected, it would strongly support the Bohm interpretation of quantum mechanics, in particular the notion of a quantum potential and the existence of a local electron trajectory. It would also unveil (and therefore confirm the existence of) the hidden variable. This in turn would imply that our conventional understanding of the uncertainty principle has to be changed. If such a radiation is proven to be non-existing, then the standard Copenhagen interpretation would sustain and the Bohm theory, one of the better candidates for the hidden variable theories \cite{Bell1987}, would not hold in its present form.

The basic proposal of Bohm's causal interpretation of quantum mechanics is that, in addition to the conventional classical potential, $V$, a particle moves according to a quantum potential, $Q$, given by
\begin{equation}
Q=-\frac{\hbar^2}{2m}\frac{\nabla^2 R}{R},
\end{equation}
where $\hbar$ is Planck's constant and $m$ is the mass of the particle, say, an electron; $R = |\psi |$, and $\psi$ is the quantum wave function which satisfies the Schr\"{o}dinger equation:
\begin{equation}
i\hbar\frac{\partial \psi}{\partial t}=-\frac{\hbar^2}{2m}\nabla^2\psi + V\psi .
\end{equation}
If one writes the wave function in polar form, $\psi=R\exp(iS/\hbar)$, then one obtains two equations:
\begin{equation}
\frac{\partial S}{\partial t}+\frac{(\nabla S)^2}{2m}+V+Q=0,
\end{equation}
\begin{equation}
\frac{\partial S}{\partial t}+\nabla\cdot\frac{(P\nabla S)}{m}=0,
\end{equation}
where $P=R^2=\psi^{\dag}\psi$. The theory then insists that Eq.(3) is indeed the Hamilton-Jacobi equation that governs the motion of the particle in the quantum mechanical world. This means that one may regard the electron as a particle with momentum $p=\nabla S$, which satisfies the equation of motion:
\begin{equation}
m\frac{d\vec{v}}{d t}=-\nabla(V+Q),
\end{equation}
It is clear that, in order to describe the motion of the particle, one must first construct the proper quantum potential of the system.

To illustrate the radiation induced by the motion governed by the quantum potential, we focus on the physics of the double-slit interference of electrons. Figure 1 shows the schematic diagram of such an experiment, which comprises an electron source $S_1$, two slits $A$ and $B$ located in the plane $S_2$, and a screen at $S_3$. The Cartesian and the polar coordinates are both indicated for the sake of our discussion below. The half-separation between the two slits is $a$ and the half-width of each slit is $b$. The quantum potential and the electron trajectories of this problem have been calculated by Philippidis et al. \cite{Philippidis} using the Feynman path-integral method to obtain the wave function. Following the same approach, we invoke, for convenience, a Gaussian variation of the slit transparency \cite{FeynmanHibbs}, $G(y) = \exp(-y^2/2b^2)$. Then the wave function for all possible paths going through either slit $A$ or $B$ and arriving at a point $(x,y)$ behind the slits is given by
\begin{eqnarray}\label{PathIntegral}
\begin{Bmatrix}
\psi_A \\
\psi_B \\
\end{Bmatrix}
=&& \sqrt{\frac{m}{2\pi i\hbar}}\Big[T+t+i\hbar\frac{tT}{mb^2}\Big]^{-1/2} \nonumber \\
&\times&\exp\Big\{-\frac{1}{2b^2}\frac{[a/T\mp(y\mp a)/t]^2}{(1/T+1/t)^2+\hbar^2/m^2b^4}\Big\} \nonumber \\
&\times&\exp\Big\{\frac{im}{2\hbar}\Big[\frac{X^2+a^2}{T}+\frac{x^2+(y\mp a)^2}{t} \nonumber \\
&&-\frac{(1/T+1/t)[a/T\mp (y\mp a)/t]^2}{(1/T+1/t)^2+\hbar^2/m^2b^4}\Big]\Big\}, 
\end{eqnarray}
where the lower signs correspond to $\psi_B$. $X$ is the distance and $T = X/v_X$ the traveling time between $S_1$ and $S_2$, and $x$ and $t = x/v_x$ that from $S_2$ to the point $(x,y) = (r,\theta)$.

\begin{figure}
\includegraphics[width=8.5cm]{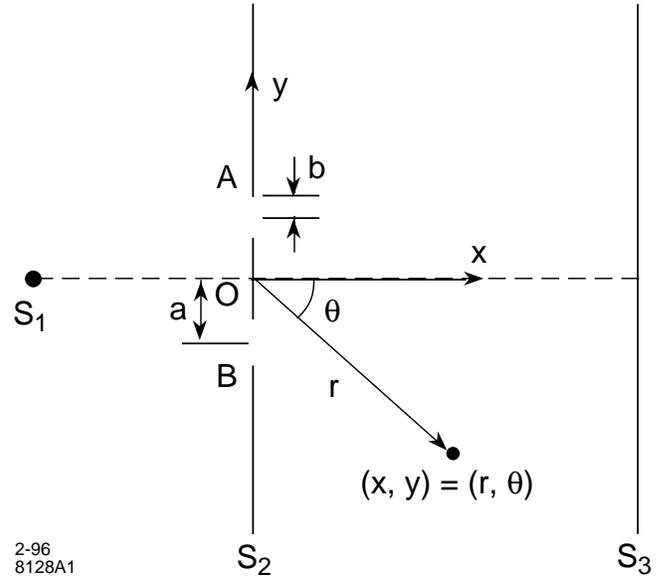}
\caption{A schematic diagram of the double-slit experiment. It comprises an electron source $S_1$, two slits $A$ and $B$ located in the plane $S_2$, and a screen at $S_3$. The half-separation between the two slits is $a$ and the half-width of each slit is $b$. 
}
\label{f1}
\end{figure}

The sum of $\psi_A$ and $\psi_B$ gives the total wave function which yields the quantum potential according to Eq.(l). A numerical calculation of the quantum potential is displayed in Fig.2. We see that one novel feature of this potential is that it consists of a few narrow and deep ÒcanyonsÓ, fanning out radially from the origin (the mid-point between the two slits). These canyons are then well separated by flat ÒplateausÓ. As demonstrated in Ref.5, an electron when coming close to a canyon will be violently accelerated and decelerated across the canyon, causing an abrupt detour of its trajectory, eventually landing on the other side of the canyon and continuing on its forward motion on the new plateau. In the end, (almost) all possible electron trajectories are confined to the plateaus. With large enough statistics, the landing positions of the electrons intercepted by a screen will show a discrete pattern, which reproduces the standard quantum mechanical interference.

To be self-consistent, our derivation should be based on a Bohm version of quantum electrodynamics, or at least a Bohm theory of quantum radiation. Not having any of these, we assume that the electron in BohmÕs theory, whose motion is governed by Eq.(5), must obey MaxwellÕs theory of electrodynamics. This assumption should be well justified since in BohmÕs theory, once the quantum potential is included, the motion of the electron is entirely causal and classical. In the case where the radiation is strong, and an emitted photon can in principle carry away a substantial fraction of the electron energy, the situation may be different in that the quantum potential after the emission will be drastically altered, and our following derivation will be invalid.
\begin{figure}
\includegraphics[width=8.5cm]{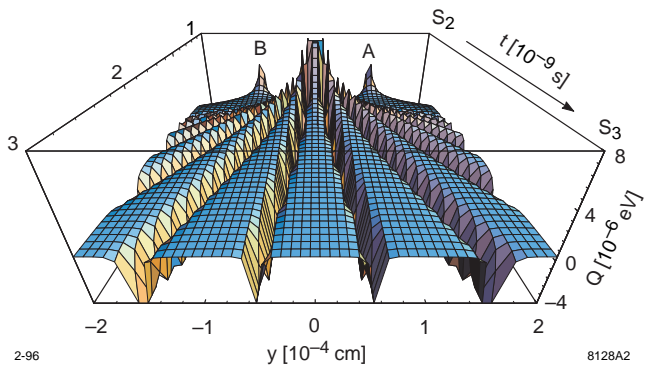}
\caption{A plot of the exact quantum potential of the double-slit system, viewed from the ÒscreenÓ located 39 cm (or $t = 3 x 10^{-9}$ s) from the slit wall. The parameters chosen are $a = 10^{-4}$ cm, $b = 1 \times 10^{-5}$ cm, $v_x= 1.3 \times 10^{10}$ cm/s, and $T = 1 \times 10^{-8}$ s.
}
\label{f2}
\end{figure}

To elucidate this radiation, it is desirable to derive analytic formulae rather than to effect a brute force numerical computation. This, however, requires a certain approximation to the very complex quantum potential shown in Fig.2. First we note that the central peak between the two slits lies in the Ògeometric shadowÓ where the electrons rarely enter. We will also neglect the two spikes at the exit of slits $A$ and $B$, which are responsible for the electron diffraction. Concentrating on the canyons and the plateaus, the following function is found to represent the quantum potential for a very wide range of all relevant parameters: $a, b, v_x=x/t\approx r/t$, and $T$:
\begin{eqnarray}
&&Q(r,\theta)\approx \nonumber \\
&&-\sum_n \frac{1}{6|n|^3}\frac{\hbar^2}{m}\frac{a^2/b^4}{1+t^2/T^2}e^{-a^2r^2(\theta-\theta_n)^2/[8b^4(1+t^2/T^2)]} , \nonumber \\
\end{eqnarray}
where $\theta_n\approx \sin\theta_n=(n\mp 1/2)\pi\hbar/mv_xa$, $n=\pm 1, \pm 2,$..., is the polar angle (from the center) of the $n$th canyon. A plot of a cross section view of our approximate quantum potential versus the exact one is shown in Fig.3; they agree very well.

\begin{figure}
\includegraphics[width=8.5cm]{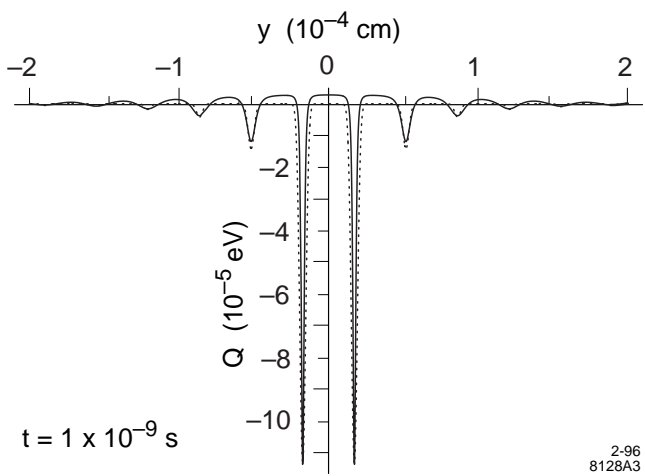}
\caption{A cross section of the exact (solid curve) and the approximate (dashed curve) quantum potential 13 cm from the screen. The parameters are the same as in Fig.2.
}
\label{f3}
\end{figure}

First we observe that the $x$ and $y$ dimensions involved in such an experiment are orders of magnitude different. Thus the electron trajectory before and after crossing the canyon is essentially parallel to the direction of the canyon, and the space-time of crossing two neighboring canyons are well separated. Therefore the radiation process can be treated as localized to each individual canyon with no need to concern about that from multiple canyon-crossings. The problem can therefore be further reduced to a single dimension transverse to the direction of the canyon (which is almost exactly the $y$-axis). Consider the $n$th canyon, and let  $y= 0$ at the center of the canyon. Then we write, with the short-hand notation $\hat{a}=a/(1+t^2/T^2)^{1/2}$,
\begin{equation}
Q_n(y)=-\frac{1}{6|n|^3}\frac{\hbar^2}{m}\frac{\hat{a}^2}{b^4} e^{-\hat{a}^2y^2/8b^4} .
\end{equation}
Since the classical potential is zero everywhere except at the wall of the double-slit, the equation of motion simply reduces to $\ddot{y}=-\nabla Q_n/m$. From Larmor's formula the radiation associated with this acceleration is
\begin{equation}
P_n=\frac{d\mathcal{E}_n}{dt}=\frac{2}{3}\frac{e^2}{c^3}|\ddot{y}|^2=\frac{2}{3}\frac{e^2}{m^2c^3}|\nabla Q_n|^2 .
\end{equation}
Now we change variable from $t$ to $y$ with $dt=y/\dot{y}$. Conservation of energy gives $\dot{y}=\sqrt{\dot{y}_{\infty}^2-2Q_n/m}\approx \sqrt{-2Q_n/m}$, where we assume the transverse velocity at infinity $\dot{y}_{\infty}=0$ for simplicity. Inserting $Q_n$ from Eq.(8), we find 
\begin{equation}
\mathcal{E}_n=\frac{\sqrt{\pi}}{81|n|^{9/2}}\frac{r_e\lambdabar_c^3 a^4/b^8}{[1+t^2/T^2]^2}mc^2,
\end{equation}
where $r_e=e^2/mc^2$ is the classical electron radius. At first look this result seems to be counter-intuitive, as the radiation is proportional to the slit separation. In fact the polar angle of the canyon $\theta_n\propto 1/a$, whereas the typical traveling time before an electron comes across a canyon is $t \propto 1/\theta_n$. Therefore $\hat{a}$ and $\mathcal{E}_n$ become asymptotically independent of $a$. Note that this radiation depends cubically on Planck's constant. This indicates that this radiation is a higher order quantum effect.

In addition to the total radiation energy, it is desirable to know its power spectrum. To do that, we start with the angular-frequency spectrum \cite{Jackson}
\begin{equation}
\frac{d^2\mathcal{E}_n}{d\omega d\Omega}=\frac{e^2\omega^2}{4\pi^2 c}\Big|\int^{\infty}_{-\infty} \dot{n}\times(\dot{n}\times \vec{\beta})e^{i\omega(t-\hat{n}\cdot\vec{r}(t)/c}dt\Big|^2 .
\end{equation}
After integrating over the solid angle, we get
\begin{equation}
\frac{d\mathcal{E}_n}{d\omega}=\frac{e^2\omega^2}{3\pi c}\Big|\int^{\infty}_{-\infty} \beta e^{i\omega t}dt\Big|^2 =\frac{e^2\omega^2}{3\pi c^3}\Big|\int^{\infty}_{-\infty} e^{i\omega t(y)}dy\Big|^2 .
\end{equation}
The problem is now reduced to finding $t$ as a function of $y$. From the relation $\dot{y}=\sqrt{-2Q_n/m}$ we have
\begin{equation}
t(y)=\sqrt{3}|n|^{3/2}\frac{m}{\hbar}\frac{b^2}{\hat{a}}\int_0^y e^{\hat{a}^2y'^2/16b^4}dy' .
\end{equation}
The exact form of this integral is too complex to render a compact result from Eq.(12). However, we note that because of the extremely nonlinear dependence of $t$ on $y$, which appears in the phase of the integral in Eq.(12), the integral rapidly diminishes beyond the first oscillation in the phase. We therefore approximate $t(y)$ to the form
\begin{equation}
t(y)\approx \frac{4|n|^{3/2}}{\sqrt{3}}\frac{m}{\hbar}\frac{b^4}{\hat{a}^2}\sinh\Big(\frac{3\hat{a}y}{4b^2}\Big).
\end{equation} 
 Inserting Eq.(14) into Eq.(12), we finally obtain the power spectrum:
\begin{equation}
\frac{d \mathcal{E}_n}{d\omega}= \frac{64}{27\pi}\frac{e^2}{c^3}\frac{b^4}{\hat{a}^2}\omega^2K_0^2(\omega\tau_n) ,
\end{equation}
where $K_0$ is the Bessel function and $r_n=(4|n|^{3/2}/\sqrt{3})(m/\hbar)b^4/\hat{a}^2$. From Eq.(15) it is easy to show that the peak of the power spectrum is located at the frequency
\begin{equation}
\omega_{max}=\frac{3}{5\tau_n}=\frac{3\sqrt{3}}{20|n|^{3/2}}\frac{\hbar}{m}\frac{a^2/b^4}{1+t^2/T^2}.
\end{equation}
Note that the depth of the quantum potential well is $|Q_n(0)|=(1/6|n|^3)(\hbar^2/m)(\hat{a}^2/b^4)\approx (0.64/|n|^{3/2})\hbar\omega_{max}$. Therefore if the electron motion was truly one-dimensional (in $y$), there would be a kinematic cut-off of the spectrum in Eq.(15) at $\hbar\omega=|Q_n(0)|$. However, in our case the major component of the momentum (and contribution to the kinetic energy) is in the $x$-dimension. So the entire spectrum is allowed. Integrating Eq.( 15) over frequency, we find
\begin{equation}
\mathcal{E}_n=\frac{\pi\sqrt{3}}{288|n|^{9/2}}\frac{r_e\lambdabar_c^3 a^4/b^8}{(1+t^2/T^2)^2}mc^2,
\end{equation}
which has the same functional form and agrees with Eq.(l0) to within $\sim 15\%$ in the numerical factor. This justifies our approximation for $t(y)$.

Now let us turn to the Copenhagen interpretation. In the classic treatment of Bloch and Nordsieck \cite{Bloch}, it was shown that in the infrared limit the mean total radiation is simply the product of the probability that the electron be scattered into an element of solid angle, and the amount of energy radiated classically in such a deflection. To make a fair comparison to our above derivation using the Bohm interpretation, in which the statistical distribution was not included, we must isolate the radiation part from the scattering probability. This means that for an electron receiving a (transverse) momentum transfer, $|\vec{q}|=q$, or equivalently a deflection angle $\theta\approx q/p_x$, the amount of radiation should be $\mathcal{E}=(2e^2/3m^2c^3)q^2/\Delta t$.

It can be easily shown (in Born approximation) that the scattering probability through the double-slit potential is $d\omega/d\theta \propto (1/q^2)\cos^2(qa/\hbar)\sin^2(qb/\hbar)$. Thus the momentum transfer associated with the first maximum is of the order $q\sim\hbar/a$. Next we fix the transit time (through the slit), $\Delta t$. We notice that in order to ensure the assumption of no transmission through the slit wall, it is necessary that the potential barrier $V_0 > E$, where $E$ is the electron kinetic energy. Since we are dealing with nonrelativistic quantum mechanics, where by definition $E < mc^2$, it is reasonable to assume $V_0 \sim mc^2$. This in turn dictates that the barrier thickness $\Delta x\gg \lambdabar_c$, if the transmission coefficient is to be kept at a negligible level. 

Putting these arguments together, we obtain the radiation energy $\mathcal{E}_1 \ll (2/3)(r_e\lambdabar_c/a^2)\beta mc^2$, where $\beta=v_x/c$. From this we conclude that although the Copenhagen interpretation also predicts a radiation in the double-slit experiment, the radiated energy has a very different functional form from that in Eq.(10). It is interesting to note that in the standard interpretation the radiation is induced in a single action, whereas in Bohm's causal interpretation the radiation due to {\it diffraction} (by the two spikes of the quantum potential at the exit of the slits, see Fig.2) and {\it interference} (by the canyons) are physically separated processes.

Consider now an experiment where $v_x=1.3\times10^{10}$ cm/sec, $a=10^{-4}$ cm, $b=10^{-6}$ cm, $t=1\times 10^{-9}$ sec, and $T=1\times 10^{-8}$ sec. In Bohm's theory the total energy radiated per crossing of a canyon, for $n = 1$, is then $\mathcal{E}_1\sim 1.7\times 10^{-8}$ eV. The radiated power is peaked at the photon energy $\hbar\omega_{max}\sim 1.9$ eV. Thus our assumption of negligible radiation reaction is satisfied. We see that the probability of emitting such a photon is  $\mathcal{E}_1/\hbar\omega_{max}\sim 1\times 10^{-8}$, which is indeed quite small. Once radiated, however, the photon can be rather energetic. In our example it is even in the visible range, i.e., $\lambda_{max}\sim 6200$ \AA. Since the maximum accelerated velocity is $\dot{y}_m/c=\sqrt{-2Q_1(0)/mc^2v}=(1/\sqrt{3})\lambda_c a)/b^2\sim 2\times 10^{-3}\ll 1$, the radiation is dipole in nature. This means the photons are emitted predominantly along the direction of the canyons, and with a good fraction land at the locations on the screen where the electrons are forbidden. Thus the photon pattern is spatially complementary to the interference pattern of the electrons. As another application, we consider the scattering of electrons off diatomic molecules. From Babinet's principle, and treating the screened atoms as black disks, one should expect a similar interference, and therefore radiation, effect. Assuming $a\sim b \sim 1$ \AA, we find that the radiation energy and frequency are comparable to the above double-slit example.

Although we investigated the double-slit experiment as an illustration, this radiation should in principle exist in any quantum mechanical system where there is a drastic variation of the quantum potential. One of the most profound implications of the EPR experiment [9] performed by Aspect et al. [10], which confirms BellÕs inequality [11], is that nonlocality is indeed a salient feature of quantum mechanics. In BohmÕs theory the nonlocality is manifested by the quantum potential. In fact, this notion has been invoked [3] to offer a dynamical explanation of the EPR paradox. Nevertheless, the conventional setting of the EPR experiment does not provide information to further distinguish the Copenhagen interpretation from that of Bohm. In this regard, our type of radiation, which should in principle also exist in the EPR experiment if the Bohm theory is correct, would be another possible test between the two interpretations. In conclusion, it is hoped that this new mechanism for radiation might help to shed some new light on one of the most intriguing and fundamental issues in physics.


{\bf Acknowledgement}

I warmly thank David Fryberger of SLAC, who introduced me to BohmÕs theory of quantum mechanics and stimulated me to carry out the investigation which led to the results of this paper. I would like to dedicate this paper to the memory of John S. Bell, whose friendship and inspiration are forever cherished. This work is supported by Department of Energy contract DE-AC02-76SF00515.

\end{document}